\documentclass[journal]{IEEEtran}
\ifCLASSINFOpdf
   \usepackage[pdftex]{graphicx}
\else
\fi
\usepackage{amsmath}
\begin{document}
\title{Noise Investigation of a Dual-Frequency VECSEL for Application to Cesium Clocks}

\author{Hui Liu,
        Gr\'egory Gredat,
        Ghaya Baili, Fran\c cois Gutty, Fabienne Goldfarb,
        Isabelle Sagnes,
        and~Fabien~Bretenaker,~\IEEEmembership{Member,~IEEE,}% <-this % stops a space
\thanks{H. Liu, G. Gredat, F. Goldfarb, and F. Bretenaker are with Laboratoire Aim\'e Cotton, CNRS, Universit\'e Paris-Sud, ENS Paris-Saclay, Universit\'e Paris-Saclay, Orsay, France.}% <-this % stops a space
\thanks{G. Baili and F. Gutty are with Thales Research \& Technology.}% <-this % stops a space
\thanks{I. Sagnes is with Centre de Nanosciences et Nanotechnologies, CNRS, Marcoussis, France.}
\thanks{Manuscript received April 19, 2005; revised August 26, 2015.}}


\markboth{Journal of \LaTeX\ Class Files,~Vol.~14, No.~8, August~2015}%
{Shell \MakeLowercase{\textit{et al.}}: Bare Demo of IEEEtran.cls for IEEE Journals}



\maketitle

\begin{abstract}
We  theoretically and experimentally study the noise of a class-A dual-frequency vertical external cavity surface emitting laser operating at Cesium clock wavelength. The intensity noises of the two orthogonally polarized modes and the phase noise of their beatnote are investigated. The intensity noises of the two modes and their correlations are well predicted by a theory based on coupled rate equations. The phase noise of the beatnote is well described by considering both thermal effects and the effect of phase-amplitude coupling. The good agreement between theory and experiment indicates possible ways to further decrease the laser noises.
\end{abstract}


% \fi
\IEEEpeerreviewmaketitle

\section{Introduction}
%\IEEEPARstart{T}{his} demo file is intended to serve as a ``starter file''
%for IEEE journal papers produced under \LaTeX\ using
%IEEEtran.cls version 1.8b and later.
%% You must have at least 2 lines in the paragraph with the drop letter
%% (should never be an issue)
%I wish you the best of success.
\IEEEPARstart{G}{enerating} low noise tunable optically-carried microwave signals is one of the core issues in the area of microwave photonics, which covers many applications such as, for instance, carrying and distributing radio signals over fibers, signal processing, antenna beam forming, distributing time and frequency standards, and so on \cite{Seed2006,Capmany2007, Yao2009}. Several techniques have been developed and extensively studied to generate such an optically-carried RF signal. Most of these techniques, such as, for example, the combination of direct modulation and optical injection locking \cite{Goldberg1983}, external modulation \cite{OReilly1992}, optical phase locking of two independent lasers \cite{Harrison1989,Fan1997}, rely on an external RF source, which is required to exhibit a low noise. This is in contrast with another technique, namely the dual-frequency laser \cite{Pillet2008}. Since the dual-frequency laser system can be simpler and cheaper, due to the absence of a low-noise external RF source, it is attracting more and more attention. The basic idea behind this two-frequency laser is that since the two laser modes share the same cavity, their optical phase noises should be essentially similar and should thus be cancelled out when one uses their beatnote as an RF signal.

So far, several schemes have been demonstrated to sustain oscillation of two laser modes in the same cavity. One scheme is based on a fiber ring laser combined with a dual-transmission-band fiber Bragg grating filter \cite{Chen2005,Xiangfei2006}. With this configuration, a beatnote at several discrete frequencies in the tens of GHz range was reported with a spectral width equal to 80 kHz . However, this laser does not offer a continuous tunability of the beatnote frequency. The second scheme is based on a solid-state laser containing a birefringent crystal (BC) to create two different optical paths for two orthogonal polarization modes inside a single cavity \cite{Alouini2001}. With this configuration, the beatnote was adjustable up to  2.7~THz \cite{Alouini1998}, and the linewidth could be less than 10~kHz. However, such solid-state lasers exhibit strong relaxation oscillations that degrade the spectral purity of the beatnote. These relaxation oscillations come from the class-B dynamical behavior of these lasers: the population inversion lifetime is typically in the 10~ps to ns range, which is not negligibly short compared to the cavity photon lifetime. This is no longer the case in a third architecture, based on a  Vertical External Cavity Surface Emitting Laser (VECSEL) containing a BC. This kind of laser is referred to as dual-frequency VECSEL (DF-VECSEL). With a high reflectivity output coupler and a few centimeter long cavity, the photon lifetime can be much longer than the lifetime of the carriers in the quantum wells, which is the condition for class-A dynamical behavior \cite{Baili2009a}. Such a DF-VECSEL has then been shown to be free from relaxation oscillations and to exhibit very low noise when oscillating in class-A regime \cite{Baili2006,Baili2007,Baili2008}. 

The first DF-VECSEL was demonstrated at 1~$\mu$m wavelength \cite{Baili2009b}, illustrating its advantages in terms of high spectral purity and large continuous beatnote frequency tunability. The same concept was then transferred at 1.55~$\mu$m wavelength \cite{De2014a} to match the telecom window, in particular for microwave photonics applications. The noises of these DF-VECSELs at 1~$\mu$m and 1.55~$\mu$m were investigated in detail. Theoretically, the intensity noises of the two modes and their correlations could be derived from a set of coupled rate equations \cite{De2013} and the phase noise of the beatnote could be accounted for by the combination of two effects: the thermal effect induced by the pump noise and  the transfer of intensity to phase noise induced by the large Henry phase/amplitude coupling effect in semiconductor VECSEL structures \cite{De2014b,De2015}. This simple model was shown to derive from a more complete theory taking spin-flipping effects into account \cite{De2014c}. In this approach, the two modes were supposed to be exactly symmetric in terms of gain and losses, allowing to expand all noises in terms of in-phase and anti-phase noise modes. The asymmetry between the pump powers and losses of the two modes was neglected. Meanwhile, the attention of researchers working on coherent population trapping (CPT) atomic clocks was also attracted by DF-VECSELs. This led to the development of a DF-VECSEL at 852 nm corresponding to the wavelength of Cesium atomic clocks \cite{Camargo2012,Dumont2014}, in view of future developments of miniature CPT atomic clocks. Nevertheless, the noise mechanisms in these lasers have not been studied in detail. Since the CPT atomic clocks have stringent requirements on the noise behavior, it is worth studying these fluctuations in more detail to have a better understanding and control of the noise.

In this paper, we thus report an experimental and theoretical investigation of the noise of a 852~nm DF-VECSEL. In Section II, we describe the implementation of this DF-VECSEL. In Section III, we present the theory of the intensity noise, the phase noise of the beatnote, and the related correlations. It is based on coupled rate equations, taking thermal effects and Henry's phase/amplitude coupling effect. While this theory has initially been developed to understand DF-VECSELs operating at telecom wavelengths, we generalized it with respect to Ref. \cite{De2015} by taking into account the difference between the pump powers that feed the two modes. In Section IV, we describe the measurement of the pump noise, intensity noises, beatnote phase noise, and the correlations between the intensity noises. All the experimental results are compared with the theoretical results. In section V, we discuss the limitations of the beatnote phase noise. This investigation thus presents a complete picture of the intensity noise and beatnote phase noise in such DF-VECSELs. Also, this work aims at testing the generality of the precedingly developed theory by applying it to the DF-VECSEL oscillating at the Cesium clock wavelength.

\section{DF-VECSEL Implementation}
\begin{figure}[]
\centering
\includegraphics[width=\columnwidth]{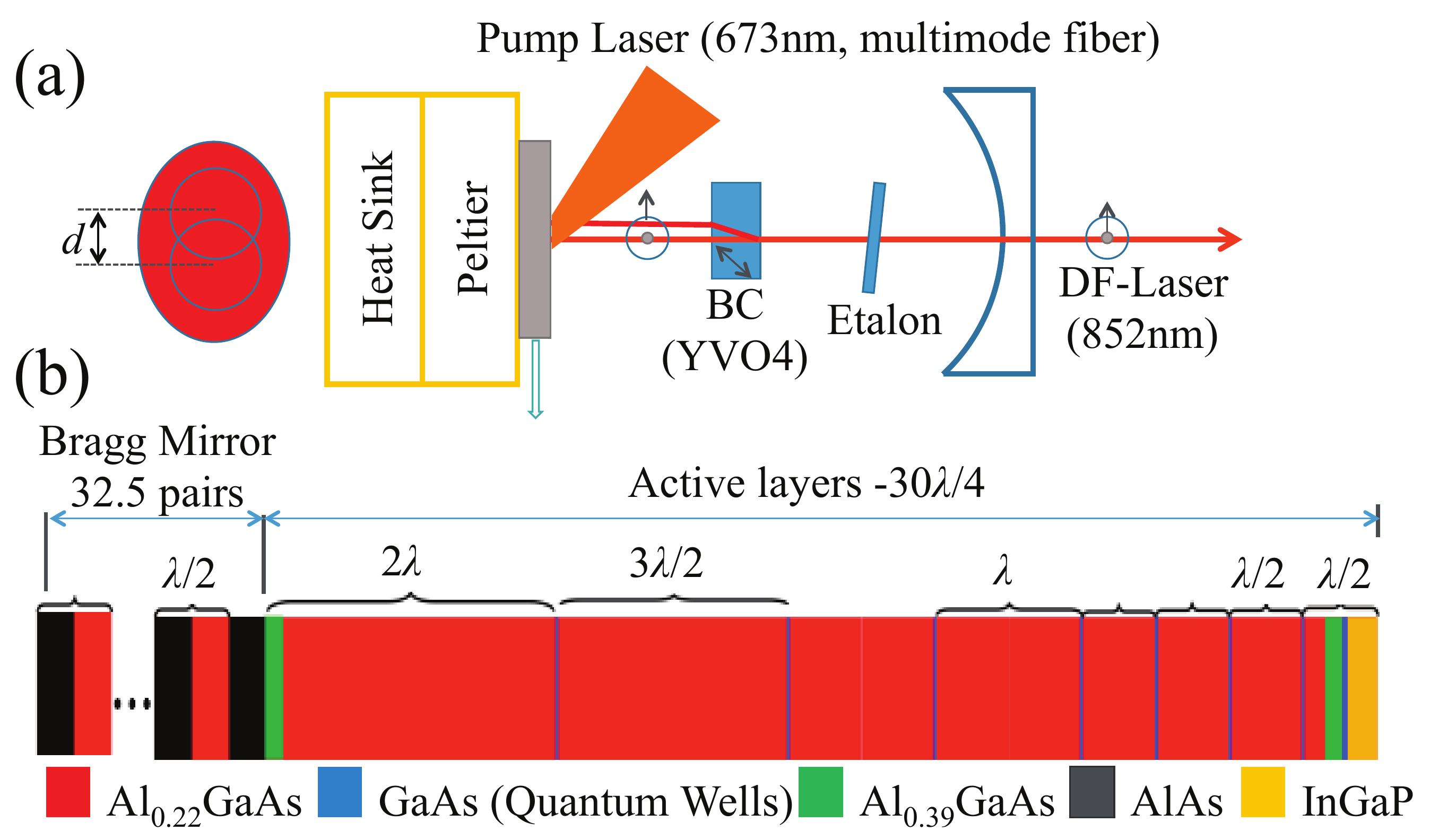}
\caption{(a) DF-VECSEL architecture, including (b) the structure of the semiconductor chip.}
\label{Fig01}
\end{figure}

The DF-VECSEL is schematized in Fig. \ref{Fig01}. The semiconductor chip is glued to a Peltier cooler, which is bonded to a heat sink. The Peltier temperature is stabilized at 20 $^{\circ}$C. The semiconductor chip is a multi-layered structure grown on a 350-$\mu$m-thick GaAs substrate by metal-organic chemical-vapor deposition method. It contains a distributed Bragg reflector and active layers. The Bragg reflector is composed of 32.5 pairs of AlAs/Al$_{0.22}$GaAs quarter-wave layers leading to a reflectivity larger than 99.94\% around 850~nm. In the active layers, seven 8-nm-thick GaAs quantum wells are embedded in Al$_{0.22}$GaAs barriers, which absorb the pump power. About 75\% of the pump power can be absorbed in single pass. Each of the quantum wells is located at an antinode of the laser field. Two layers of Al$_{0.39}$GaAs form potential barriers to confine the carriers. A 50-nm-thick InGaP and 5~nm GaAs layer cap the structure to prevent the Al oxidation. This chip is designed without anti-reflection coating to increase the gain of the mode resonant within the micro-cavity created in the semiconductor structure.
	
An anti-reflection coated YVO$_4$ birefringent crystal (BC) is inserted inside the cavity. It is cut at 45$^{\circ}$ of its optical axis, thus separating the extraordinary polarization from the ordinary one by a distance denoted as $d$. Here, the thickness of the BC is 0.5~mm, leading to a value of $d$ close to 50~$\mu$m. This separation leads to the oscillation of two orthogonally polarized laser modes inside the cavity, whose centers are distant from each other by $d$ in the active structure. However, depending on the mode radius $w_0$ in the structure, the two modes partially overlap inside the gain structure and thus experience some competition through nonlinear coupling, i. e., gain cross-saturation \cite{Pal2010,Sargent1975}. An etalon is also introduced to improve the robustness of single mode oscillation for each of the two orthogonal polarizations. The laser output coupler is a concave mirror with a transmission of 0.5\% and a radius of curvature of 5~cm. If we consider the losses of the output coupler, the photon lifetime inside the 5-cm-long cavity is 32~ns, which is much longer than the carrier lifetime inside the wells, thus ensuring that the laser obeys class-A dynamics. The pump laser is a 673~nm laser diode. It is delivered to the semiconductor chip by a multimode fibre whose core diameter is 102~$\mu$m with a numerical aperture equal to 0.22. After collimation, the pump beam is incident on the structure with an angle of about 40$^{\circ}$.
	
\begin{figure}[]
\centering
\includegraphics[width=0.75\columnwidth]{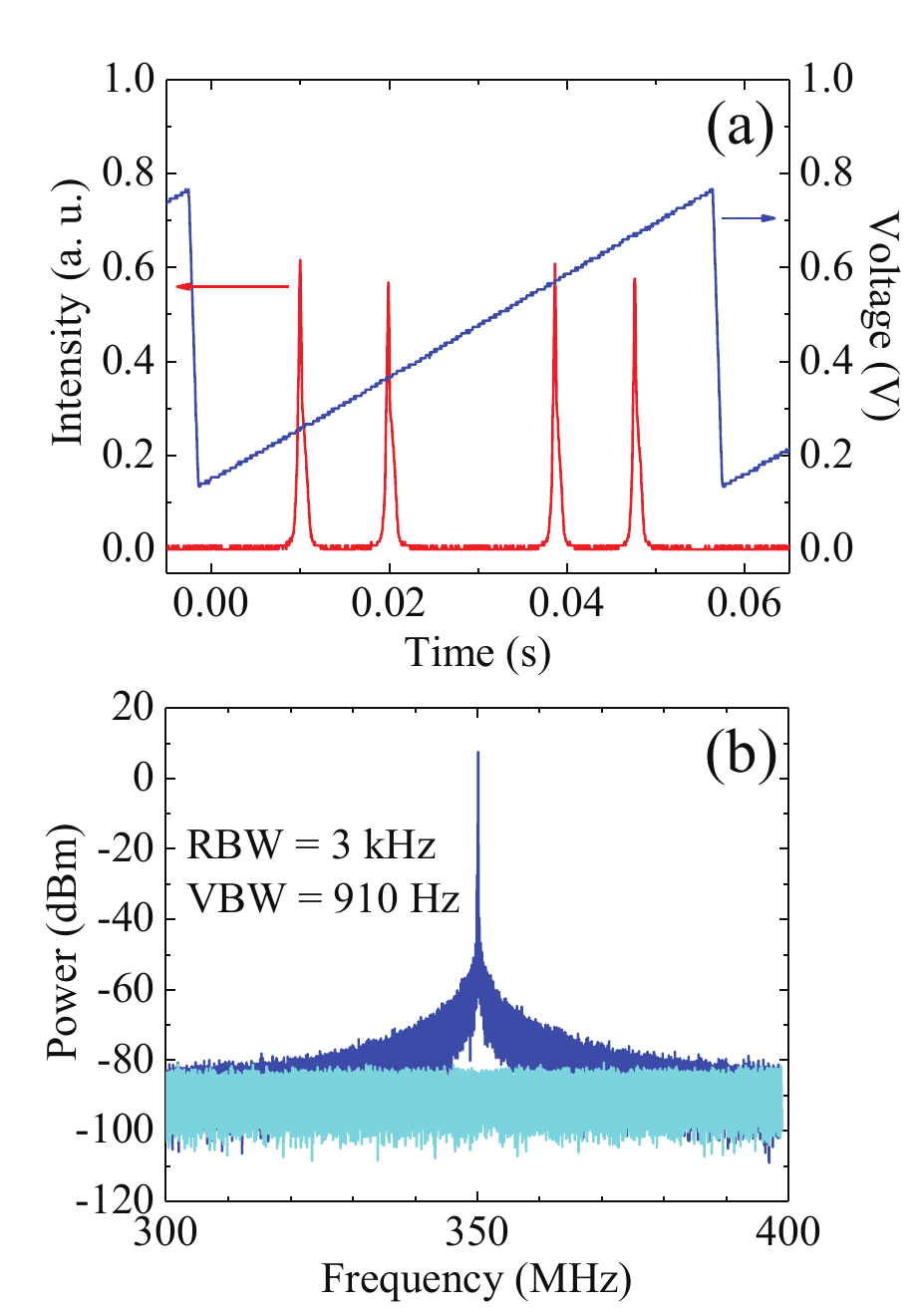}
\caption{(a) Spectrum of the DF-VECSEL visualized thanks to a Fabry-Perot interferometer, with a free spectral range equal to 10~GHz. The ramp applied to the Fabry-Perot is also shown. (b) Beatnote spectrum obtained using an electrical spectrum analyzer. The light horizontal trace is the measurement noise floor.}
\label{Fig02}
\end{figure}

	The two orthogonally polarized laser modes are monitored by a Fabry-Perot interferometer (FPI), as shown in Fig.\;\ref{Fig02}(a) for a beatnote frequency of the order of 3~GHz. The spectrum of the beatnote, measured by an electric spectrum analyzer, is shown in Fig. \ref{Fig02}(b). Due to the 1~GHz bandwith of the detector, we adjusted the frequency of the beatnote at 350~MHz. As can be seen in this spectrum, there is a strong noise pedestal, extending up to several tens of MHz, at the foot of the beatnote peak. This pedestal results mainly from the beatnote phase noise, which will be investigated in the following sections.

\section{Theory of the Intensity and Beatnote Phase Noises}
\subsection{Intensity Noises}
The laser noise can be derived with the rate equations formalism used in Ref. \cite{De2013}. The two modes are described by their intracavity photon numbers $F_{0x}$ and $F_{0y}$, and the pump creates two carrier reservoirs for the two modes. The corresponding unsaturated carrier numbers are noted $N_{0x}$ and $N_{0y}$. Some nonlinear coupling is introduced between the rate equations governing the two modes, in order to describe gain cross-saturation \cite{De2013}, through the ratios $\xi_{xy}$ and $\xi_{yx}$ of the cross- to self-saturation coefficients. The steady-state numbers of photons $F_{0x}$ and $F_{0y}$  in the $x$- and $y$-polarized modes are given by \cite{De2013}:
\begin{align}
F_{0x}&=\frac{1}{\kappa\tau}\frac{(r_x-1)-\xi_{xy}(r_y-1)}{1-C}\ ,\label{eq01}\\
F_{0y}&=\frac{1}{\kappa\tau}\frac{(r_y-1)-\xi_{yx}(r_x-1)}{1-C}\ ,\label{eq02}
\end{align}
where $\tau$ is the carrier lifetime, $r_x$ and $r_y$ are the excitation ratios of the two modes, $\kappa$ is the coupling coefficient between the photons and the carriers, which is proportional to the laser cross section of the transition. $C$ is the coupling constant between the two modes defined by
\begin{equation}
C=\xi_{xy}\xi_{yx}\ .\label{eq03}
\end{equation}
The steady-state solution of Eqs. (\ref{eq01}) and (\ref{eq02}) is stable only when $C<1$. 

The pump fluctuations are introduced as fluctuations of the unsaturated numbers of carriers of the two modes:
\begin{equation}
N_{0i}(t)=\overline{N}_{0i}+\delta N_{0i}(t)\ \label{eq03N1}
\end{equation}
for $i=x,y$ where the average values are related to the excitation ratios through:
\begin{equation}
\overline{N}_{0i}=\frac{r_i}{\kappa\tau_i}\ .\label{eq03N2}
\end{equation}
If we call $\delta F_x(t)$ and $\delta F_y(t)$ the (small) fluctuations of the number of photons of the two modes, Ref. \cite{De2013} shows that their Fourier transforms are related to the Fourier transforms of $\delta N_{0x}(t)$ and $\delta N_{0y}(t)$ according to:
\begin{equation}
\left[\begin{array}{c}
\widetilde{\delta F_x}(f)\\
\widetilde{\delta F_y}(f)
\end{array}
\right]=\left[\begin{array}{cc}M_{xx}(f)&M_{xy}(f)\\M_{yx}(f)&M_{yy}(f)\end{array}\right]\left[\begin{array}{c}
\widetilde{\delta N_{0x}}(f)\\
\widetilde{\delta N_{0y}}(f)
\end{array}
\right]
\ ,\label{eq04}
\end{equation}
where tilde denotes Fourier transformed quantities and $f$ is the noise frequency. The matrix elements are given by:
\begin{align}
&M_{xx}(f)=\frac{1}{\Delta(f)\tau}\left[\frac{1}{\tau_y} -2\mathrm{i}\pi f\frac{r_y/\tau-2\mathrm{i}\pi f}{\kappa F_{y0}}\right]\ ,\label{eq05}\\
&M_{xy}(f)=-\frac{\xi_{xy}}{\tau\tau_x\Delta(f)}\ ,\label{eq06}
\end{align}
with similar expressions for $M_{yy}$ and $M_{yx}$. $\tau_x$ and $\tau_y$ are lifetimes of the photons for the two modes and $\Delta$ is given by 
\begin{align}
\Delta(f)=&\left[\frac{1}{\tau_x} -2\mathrm{i}\pi f\frac{r_x/\tau-2\mathrm{i}\pi f}{\kappa F_{x0}}\right]\nonumber\\&\times\left[\frac{1}{\tau_y} -2\mathrm{i}\pi f\frac{r_y/\tau-2\mathrm{i}\pi f}{\kappa F_{y0}}\right]-\frac{C}{\tau_x\tau_y}\ .\label{eq07}
\end{align}

In the following we suppose that, even if the two modes see different pump powers $P_{\mathrm{p},x}$ and $P_{\mathrm{p},y}$, the relative intensity noises (RINs) of the two pumps are the same, defined as
\begin{equation}
\mathrm{RIN}_{\mathrm{p}}(f)=\frac{\langle|\widetilde{\delta N_{0x}}(f)|^2\rangle}{\overline{N}_{0x}^2}=\frac{\langle|\widetilde{\delta N_{0y}}(f)|^2\rangle}{\overline{N}_{0y}^2}\ ,\label{eq09}
\end{equation}
%In the simple case where we suppose that the pump powers and pump noise power spectral densities are the same for the two modes, namely
%\begin{equation}
%\overline{N}_{0x}=\overline{N}_{0y}\equiv N_0\ ,\label{eq08}
%\end{equation}
%and 
%\begin{equation}
%\langle|\delta N_{0x}(f)|^2\rangle=\langle|\delta N_{0y}(f)|^2\rangle\equiv\mathrm{RIN}_{\mathrm{p}}(f)N_0^2\ ,\label{eq09}
%\end{equation}
where $\langle\quad\rangle$ denotes statistical average. Eq. (\ref{eq04}) leads to the following expressions for the RIN of the two modes:
%\begin{equation}
%\mathrm{RIN}_i(f)=\left[\frac{|M_{ii}|^2r_i^2/\tau_i^2+|M_{ii}|^2r_j^2/\tau_j^2+\eta\left(M_{ii}M_{ij}^*+M_{ii}^*M_{ij}\right)}{\kappa^2 F_{0x}^2}\right]\mathrm{RIN}_{\mathrm{p}}(f)
%\end{equation}
\begin{align}
\mathrm{RIN}_i(f)&=\frac{\left|\widetilde{\delta F_i}(f)\right|^2}{F_{0i}^2}=\left[\frac{|M_{ii}|^2r_i^2}{\tau_i^2}+\frac{|M_{ij}|^2r_j^2}{\tau_j^2}\right.\nonumber\\
&+\left.\eta\frac{r_ir_j}{\tau_i\tau_j}\left(M_{ii}M_{ij}^*+M_{ii}^*M_{ij}\right)\right]\frac{\mathrm{RIN}_{\mathrm{p}}(f)}{\kappa^2 F_{0i}^2}\ ,\label{eq10}
\end{align}
where $i,j=x,y$ with $i\neq j$. In this expression, $\eta$ is the modulus of the correlation between the two pump noises defined by
\begin{equation}
\langle\widetilde{\delta N_{0x}}(f)\widetilde{\delta N_{0y}}^*(f)\rangle=\eta\,\mathrm{RIN}_{\mathrm{p}}(f)\overline{N}_{0x}\overline{N}_{0y}\mathrm{e}^{\mathrm{i}\Psi}\ ,\label{eq11}
\end{equation}
where $\Psi$ is the phase of this correlation. In the following, we suppose that both $\eta$ and $\Psi$ do not depend on the noise frequency $f$, a hypothesis that will be justified by the measurements reported below.
\subsection{Beatnote Phase Noise}
The phase noise of the beatnote has been shown to originate from two main contributions \cite{De2014b}: i) the fluctuations of the phase induced by the strong phase/intensity coupling in the semiconductor and ii) the fluctuations of the temperature of the semiconductor chip induced by the fluctuations of the pump power. The first contribution leads to a beatnote phase noise $\delta\phi_{\mathrm{H}}(t)$ whose Fourier transform is given by:
\begin{equation}
\widetilde{\delta\phi}_{\mathrm{H}}(f)=\frac{\alpha}{2}\left(\frac{\widetilde{\delta F_x}(f)}{F_{0x}}-\frac{\widetilde{\delta F_y}(f)}{F_{0y}}\right)\ ,\label{eq12}
\end{equation}
where $\alpha$ is Henry's factor. The thermal effect leads to the following phase noise for the beatnote:
\begin{equation}
\widetilde{\delta\phi}_{\mathrm{T}}(f)=-\frac{\omega_0 \Gamma_{\mathrm{T}}R_{\mathrm{T}}}{2\mathrm{i}\pi f(1+2\mathrm{i}\pi f\tau_{\mathrm{T}})}\left[P_{\mathrm{p},x}\frac{\widetilde{\delta N_{0x}}}{\overline{N}_{0x}}-P_{\mathrm{p},y}\frac{\widetilde{\delta N_{0y}}}{\overline{N}_{0y}}\right]\ ,\label{eq13}
\end{equation}
where $\omega_0$ is the laser angular frequency, $R_{\mathrm{T}}$ and $\tau_{\mathrm{T}}$ are the thermal resistance and response time of the semiconductor structure, and $\Gamma_{\mathrm{T}}$  is the refractive index variation with temperature, given by
\begin{equation}
\Gamma_{\mathrm{T}}=\frac{L_{\mathrm{SC}}}{L_{\mathrm{ext}}}\frac{d\overline{n}}{dT}\ ,\label{eq14}
\end{equation}
where $d\overline{n}/dT$ is the temperature dependence of the average refractive index in the structure and $L_{\mathrm{SC}}$ and $L_{\mathrm{ext}}$ are the lengths of the semiconductor structure and the external cavity, respectively.

Since the two contributions given by Eqs. (\ref{eq12}) and (\ref{eq13}) originate from the same source of noise, i. e. the pump noise, they must be added coherently, leading to the following expression for the beatnote phase noise power spectral density:
\begin{equation}
\left|\widetilde{\delta\phi}_{\mathrm{H+T}}(f)\right|^2=\left[|Q_x|^2+|Q_y|^2+2\eta\mathrm{Re}(Q_xQ_y^*\mathrm{e}^{\mathrm{i}\Psi})\right]\mathrm{RIN}_{\mathrm{p}}(f)\ ,\label{eq15}
\end{equation}
with
\begin{align}
Q_x(f)&=\frac{\alpha r_x}{2\kappa\tau_x}\left(\frac{M_{xx}}{F_{0x}}-\frac{M_{yx}}{F_{0y}}\right)-\frac{\omega_0 \Gamma_{\mathrm{T}}R_{\mathrm{T}}}{2\mathrm{i}\pi f(1+2\mathrm{i}\pi f\tau_{\mathrm{T}})}P_{\mathrm{p},x}\ ,\label{eq16}\\
Q_y(f)&=\frac{\alpha r_y}{2\kappa\tau_y}\left(\frac{M_{yy}}{F_{0y}}-\frac{M_{xy}}{F_{0x}}\right)+\frac{\omega_0 \Gamma_{\mathrm{T}}R_{\mathrm{T}}}{2\mathrm{i}\pi f(1+2\mathrm{i}\pi f\tau_{\mathrm{T}})}P_{\mathrm{p},y}\ .\label{eq16N1}
\end{align}
%for $i,j=x,y$ and $i\neq j$.
\subsection{Noise Spectral Correlations}
Since the intensity noises and the phase noise mainly originate from the pump noise, one can expect that there should be some correlations between these noises. These correlations can be helpful to have a deeper insight of the physics of this laser. In the following, we will measure some of these correlations and their frequency dependence. For example, we give here the definition of the spectral correlation between the intensity noises of the two modes:
\begin{equation}
\Theta_{\delta F_x-\delta F_y}(f)=\frac{\left\langle\widetilde{\delta F_x}(f)\widetilde{\delta F_y^*}(f)\right\rangle}{\sqrt{\left\langle\left|\widetilde{\delta F_x}(f)\right|^2\right\rangle\left\langle\left|\widetilde{\delta F_y}(f)\right|^2\right\rangle}}\ .\label{eq17}
\end{equation}
Similar expressions can be introduced for the correlations between any of the intensity noises $\delta F_x$ and $\delta F_y$ and the beatnote phase noise $\delta\phi$.

\section{Noise Measurements and Modeling}
\subsection{Pump Noise}
From the preceding section, we expect both the RIN and the beatnote phase noise to depend on the RIN and correlation of the pump noises. It is thus necessary to fully characterize the pump noise, as described in the present Section.

\begin{figure}[]
\centering
\includegraphics[width=\columnwidth]{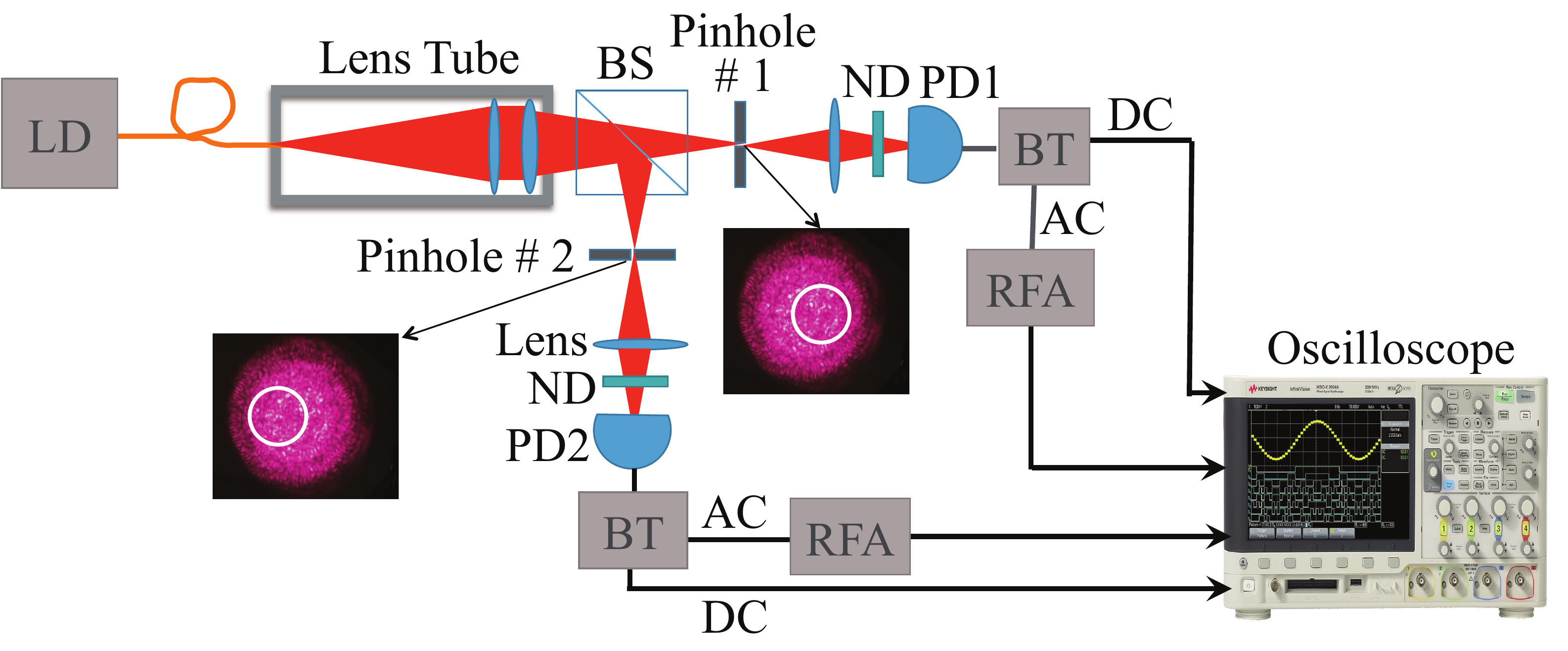}
\caption{Experimental setup used to measure the RIN of the pumps and the correlation between the pump noises. LD: pump laser diode; BT: bias tee; ND: neutral density filter; PD1, PD2: photodiodes; BS: beam splitter; RFA: radio frequency amplifier.}
\label{Fig03}
\end{figure}

The pump laser is multimode fiber coupled. Consequently, the numerous spatial modes of the fiber lead to a complicated speckle pattern at the output of the fiber. This speckle pattern will evolve when the spectral content of the laser evolves or when the path differences between the fiber modes evolve. Moreover, since the two polarization modes of the DF-VECSEL are spatially separated in the active medium and have a smaller diameter than the pump spot, the spatial regions of the pump spot that pump the two laser modes are different.  This fact requires us to extract the pump noises just in the spatial regions that intersect the two VECSEL modes. Moreover, to extract the correlations between the pump noises for the two modes, we must record the corresponding pump intensity fluctuations simultaneously. To this aim, we built the experiment setup schematized in Fig.\;\ref{Fig03}. The laser diode (LD) and the lens tube are the ones used to pump the DF-VECSEL. A beam splitter then splits the pump beam into two identical arms. At the focal point, two pinholes of 50 $\mu$m radii are placed to mimic the two laser modes. The white circles in the two pictures in Fig.\ \ref{Fig03} show the relative diameter of these pinholes with respect to the pump spot. Each of the pinholes is controlled by a three dimension translation stage. We can thus record the two pump noises as a function of their separation $d$ on the structure. After the pinholes, the beams are sent through lenses and neutral density filters before being detected by two photodiodes with 2.4~mm $\times$ 2.8~mm photosensitive areas and 30 MHz bandwidth. The photocurrent signals are separated into DC and AC signals by a bias-tee. The AC signals are amplified by RFAs. The two signals are sampled by a digital oscilloscope simultaneously. Further processing, based on Fourier transforming the pump noises signals, is performed on a computer. Averaging over many samples permits to obtain the pump RINs and correlations following Eqs.\ (\ref{eq09}) and (\ref{eq11}).

%\begin{figure}[]
%\centering
%\includegraphics[width=\columnwidth]{Fig04N1.eps}
%\caption{Pump RIN spectrum measured using the setup of Fig.\ \ref{Fig03}.}
%\label{Fig04}
%\end{figure}
\begin{figure}[h!]
\centering
\includegraphics[width=0.9\columnwidth]{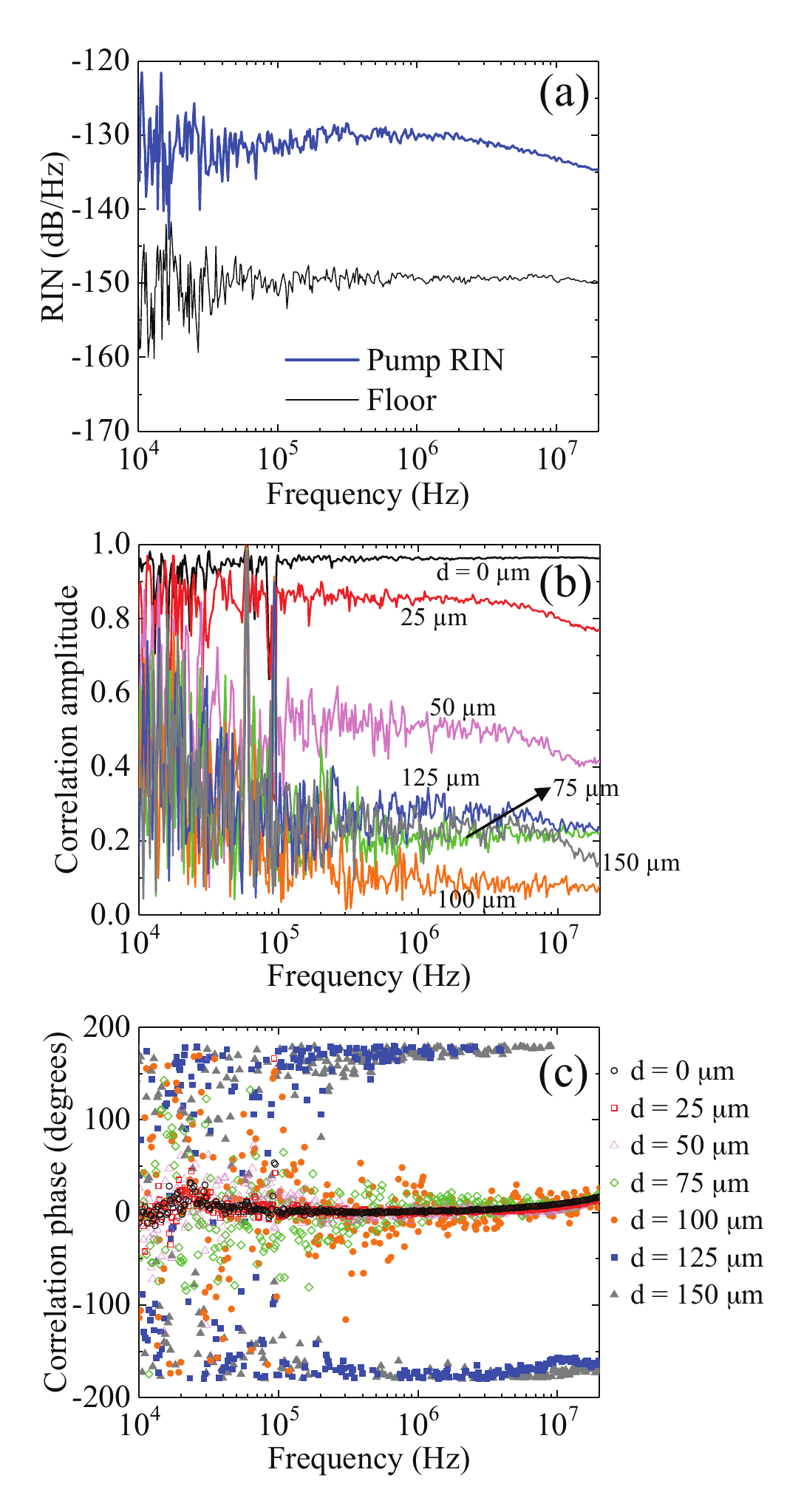}
\caption{(a) Pump RIN spectrum measured using the setup of Fig.\ \ref{Fig03}. (b) Amplitude and (c) phase of the correlation spectrum of the pump noises for different values of the mode separation $d$.}
\label{Fig04}
\end{figure}

Figure \ref{Fig04} shows the pump noise characterization results obtained with the setup of Fig.\;\ref{Fig03}.  First of all, the RIN of the pump laser is reproduced in Fig.\;\ref{Fig04}(a). For simplicity, in subsequent modeling, we take the RIN of the pump to be white in the 10 kHz to 20 MHz frequency range, and take it equal to $-130\pm 3\;\mathrm{dB/Hz}$. 
%\begin{figure}[]
%\centering
%\includegraphics[width=\columnwidth]{Fig05N1.eps}
%\caption{(a) Amplitude and (b) phase of the correlation spectrum of the pump noises for different values of the mode separation $d$.}
%\label{Fig05}
%\end{figure}
Figures \ref{Fig04}(b) and \ref{Fig04}(c) show the results of the measurements of the correlations between the two pump noises, selected by the two 50 $\mu$m radius pinholes whose images are separated by a distance $d$ in the transverse plane. The measurement starts with $d = 0$, which can be validated by making the correlation amplitude close to one, illustrating the identity of the two pump regions passing through the pinholes. When $d$ increases from 0 to 100 $\mu$m, the correlation amplitude decreases, and the correlation phase concentrates around zero. However, when $d$ increases up to 125 $\mu$m and beyond, the correlation amplitude increases again, and the correlation phase tends to $\pm180^{\circ}$. 

\subsection{RIN and Beatnote Phase Noise of the DF-VECSEL}
\begin{figure}[]
\centering
\includegraphics[width=\columnwidth]{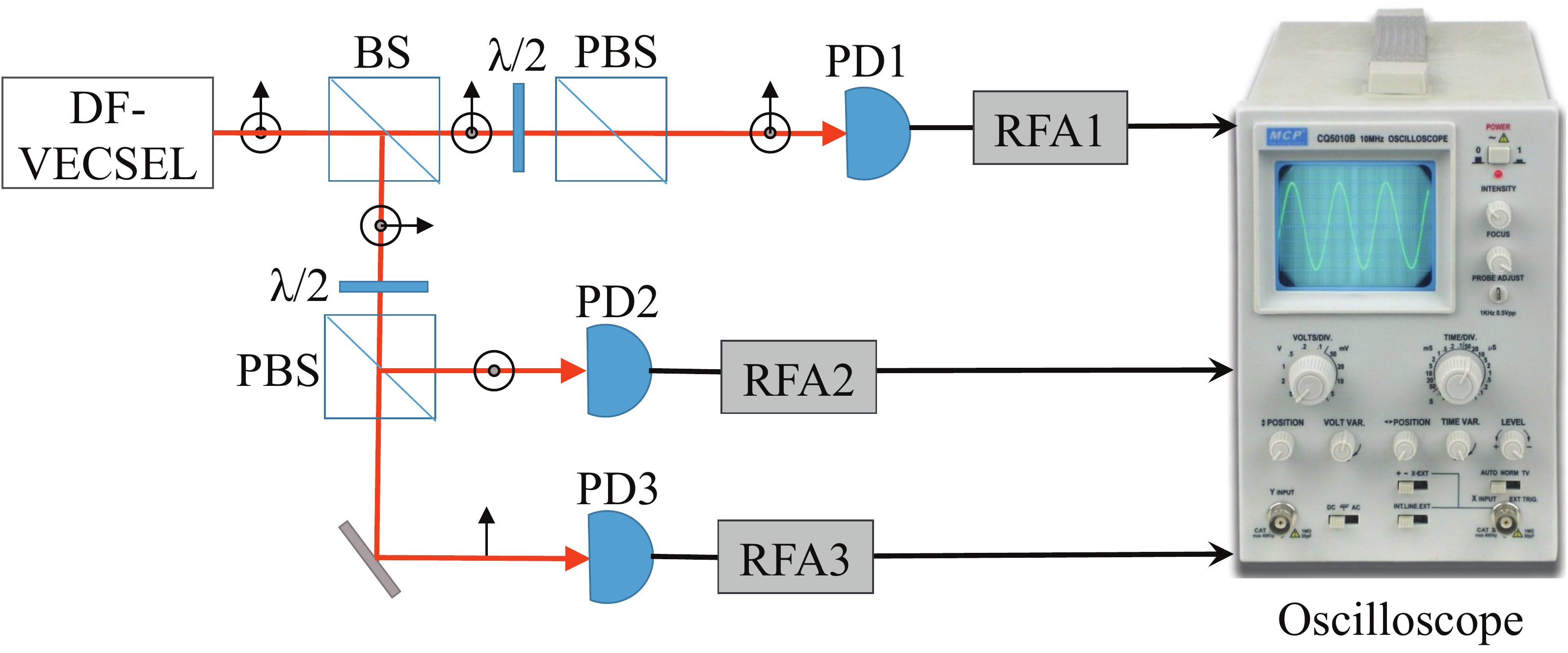}
\caption{Experimental setup used to measure the noises of the DF-VECSEL. BS: beam splitter; PBS: polarization beam splitter; PD's: photodiodes; RFA: radio frequency amplifiers.}
\label{Fig06}
\end{figure}

The intensity noises and beatnote phase noise are measured simultaneously by the experiment setup shown in Fig.\ \ref{Fig06}. The beam emitted by the DF-VECSEL is split into two arms. In order to measure the intensity noises of the two cross-polarized modes separately, the two orthogonal polarized lasers in one arm are separated by using a half-wave plate and a PBS. In the other arm, a polarizer projects the o and e modes to the same polarization direction, in order to detect the beatnote. The three beams are sent into three PDs with a 1~GHz bandwidth. After amplification, the intensity noises and the beatnote signals are simultaneously recorded by a deep memory digital oscilloscope.

In order to test the robustness of the predictions of our model, we measured all the noises for two values of the cavity length, namely 48.0 mm and 49.5 mm. These two cavity lengths correspond to two values of the beam waist $w_0=65\ \mu\mathrm{m}$ and  $40\ \mu\mathrm{m}$, respectively. With a separation $d=50\ \mu\mathrm{m}$, we take the coupling constant to be equal to $C = 0.44$ and $C=0.15$, respectively. Thus, by changing the cavity length, we change three parameters relevant to noise: the coupling constant $C$, the pump noise level and the pump noise correlation $\eta$. 

\begin{figure}[]
\centering
\includegraphics[width=\columnwidth]{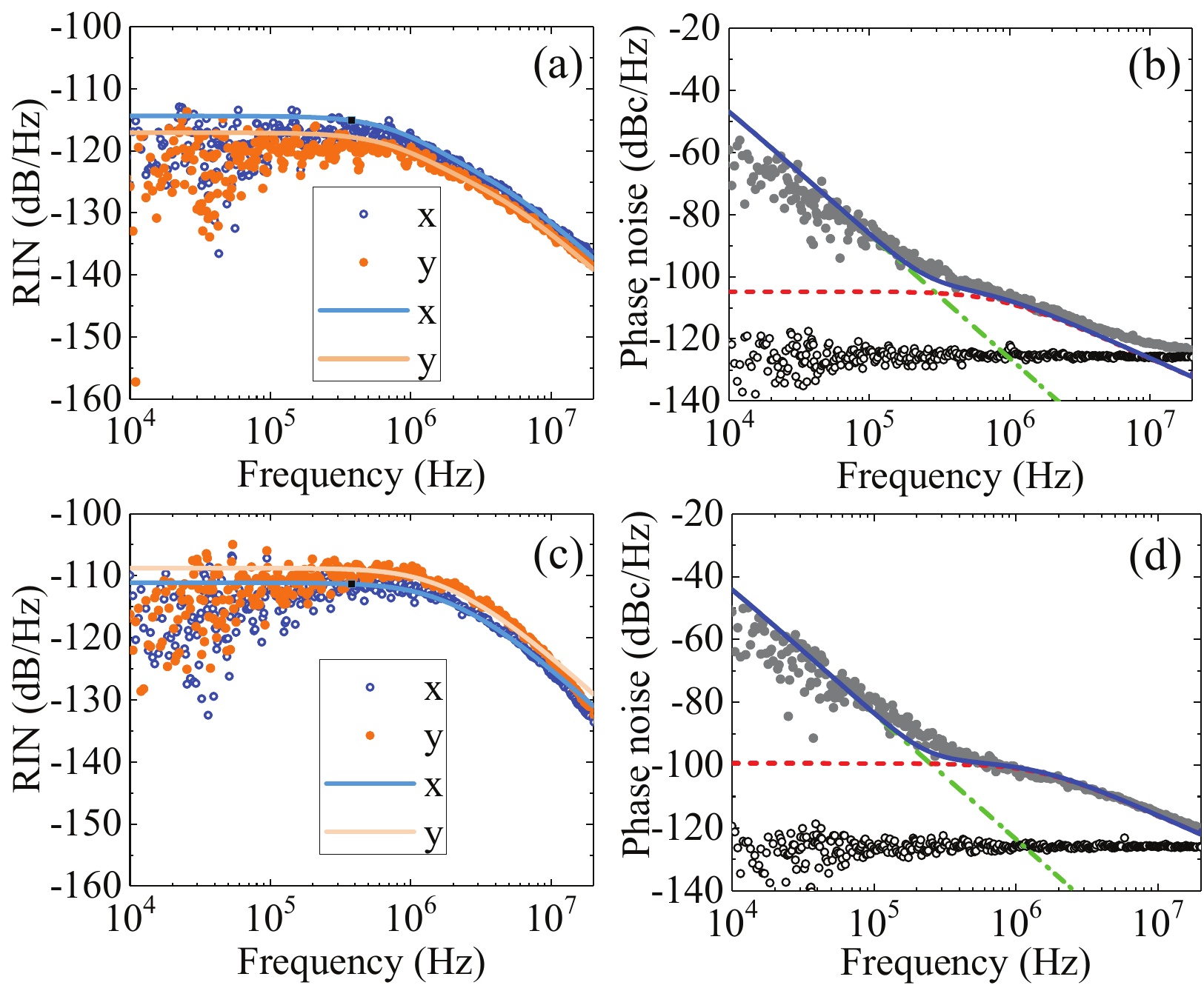}
\caption{Experimental measurements of the (a,c) intensity noises and (b,d) beatnote phase noise of the DF-VECSEL. The cavity length is 48.0 mm in (a,b) and 49.5 mm in (c,d). The dots are measurements and the full lines are the corresponding theoretical spectra computed with $\tau = 1\,\mathrm{ns}$ , $\Psi=0$, $\alpha = 5.2$, $R_{\mathrm{T}} = 40\,\mathrm{K.W}^{-1}$ and (a,b) $\tau_x = 13\,\mathrm{ns}$, $\tau_y = 16\,\mathrm{ns}$, $r_x = 1.59$, $r_y = 1.63$, $C=0.44$, $\mathrm{RIN}_{\mathrm{p}}= -133\,\mathrm{dB/Hz}$, $P_{\mathrm{p},x} = 0.42\,\mathrm{W}$, $P_{\mathrm{p},y} = 0.35\,\mathrm{W}$, $\Gamma_\mathrm{T} = 1.34 \times 10^{-7}\,\mathrm{K}^{-1}$, $\tau_\mathrm{T} = 45\,\mu\mathrm{s}$, $\eta=0.45$; (c,d)  $\tau_x = 10\,\mathrm{ns}$, $\tau_y = 8\,\mathrm{ns}$, $r_x = 1.25$, $r_y = 1.22$, $C=0.15$, $\mathrm{RIN}_{\mathrm{p}}= -129\,\mathrm{dB/Hz}$, $P_{\mathrm{p},x} = 0.27\,\mathrm{W}$, $P_{\mathrm{p},y} = 0.33\,\mathrm{W}$, $\Gamma_\mathrm{T} = 1.32 \times 10^{-7}\,\mathrm{K}^{-1}$, $\tau_\mathrm{T} = 45\,\mu\mathrm{s}$, $\eta=0.1$. In (b,d) the dashed (resp. dot-dashed) line is the contribution to the beatnote phase noise coming from amplitude phase coupling (resp. thermal effect).}
\label{Fig07}
\end{figure}
Figure\ \ref{Fig07} shows the resulting RIN and beatnote phase noise spectra, both theoretical and experimental. Figs.\ \ref{Fig07}(a) and (b) were obtained when the length of the cavity is 48mm, while (c) and (d) correspond the cavity length equal to 49.5mm. The RIN spectra of Figs.\ \ref{Fig07}(a) and (c) exhibit the typical first-order filter shape consistent with the expected class-A dynamical behavior of the laser. Comparing (a) and (c), one can notice that the spectra in (c) have a sharper corner than the curves in (a). This is due to the fact that $C$ is larger in (a), inducing a stronger deformation with respect to a pure first-order filter shape. The RINs of the two modes are slightly different because they experience slightly different gains and losses. In all cases the experimental results are well reproduced by the model. 

The phase noise spectra are shown in Figs.\ \ref{Fig07}(b) and (d). From the two figures, one can see that both spectra exhibit a change of slope at about 300 kHz. The corresponding theoretical spectra show that this is due to the fact that the thermal effect dominates at low frequencies and phase/amplitude coupling effect dominates at higher frequencies. The transition between these two effects just occurs at around 300 kHz. 

The comparison of Figs.\;\ref{Fig07}(c) and \ref{Fig07}(d) with respect to Figs.\;\ref{Fig07}(a) and \ref{Fig07}(b) shows that both the intensity and the phase noises increase when the cavity length is increased, i. e., when the mode size in the gain region decreases. This is due to the combination of three factors: i) the increase of the pump noise seen by the laser mode; ii) the decrease of the coupling constant $C$, and iii) the decrease of the pump noise correlation amplitude $\eta$. The influence of all these factors will be discussed in Section \ref{discussion}.
\subsection{Correlation Behavior}
To gain a deeper understanding of the noise and perform a more solid test of the theory, the correlation between the intensity noises of the two modes is also investigated. 

\begin{figure}[]
\centering
\includegraphics[width=\columnwidth]{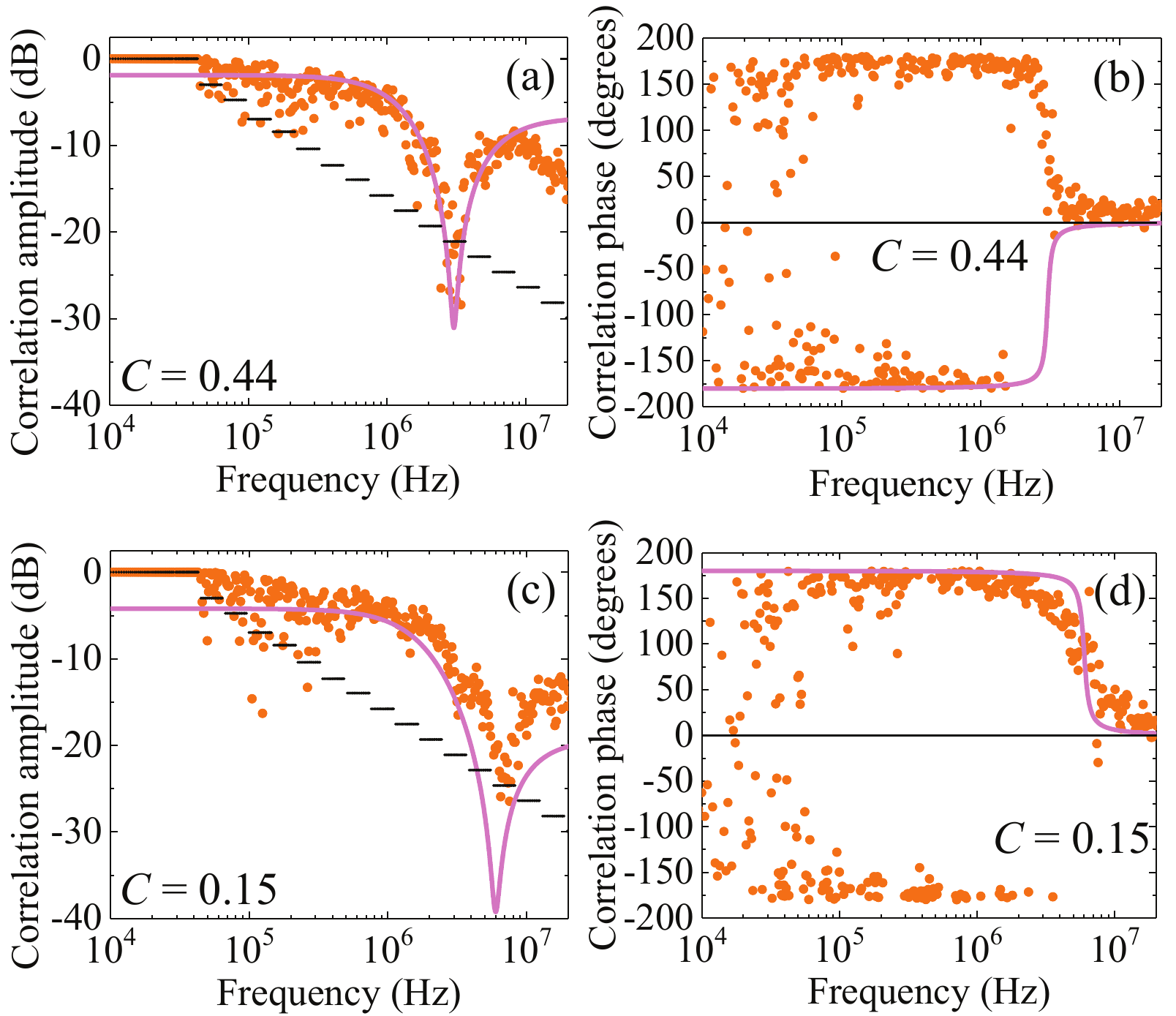}
\caption{Experimental measurements of the (a,c) amplitudes and (b,d) phases of the correlation spectra between the intensity noises of the two modes of the DF-VECSEL. The cavity length is 48.0 mm in (a,b) and 49.5 mm in (c,d). The dots are measurements and the full line are the corresponding theoretical spectra, obtained with the same parameters as in Fig.\ \ref{Fig07}.}
\label{Fig08}
\end{figure}
Figure\ \ref{Fig08} reproduces the correlation spectra between the intensity noises of the two orthogonal polarized laser modes. As can be seen, the correlation amplitude is flat and strong at  lower frequencies, before dropping and exhibiting a minimum in a narrow frequency band, after which it increases again. The corresponding correlation phase changes from about $\pm180^{\circ}$ to 0. This spectral behavior can be explained by considering the DF-VECSEL as a coupled oscillator system. Then, this system has two noise eigenmodes, which are linear combinations of the two intensity noises:
\begin{align}
\widetilde{\delta F_1}(f)&=c_{1x}\widetilde{\delta F_x}(f)+c_{1y}\widetilde{\delta F_y}(f)\ ,\label{eq18}\\
\widetilde{\delta F_2}(f)&=c_{2x}\widetilde{\delta F_x}(f)-c_{2y}\widetilde{\delta F_y}(f)\ .\label{eq19}
\end{align}
If we suppose, as an approximation, that the two laser modes have the same gains and losses, the constants $c_{1x}$, $c_{1y}$, $c_{2x}$, and $c_{2y}$ are then all be equal. Then, the two noise eigenmodes become in-phase and anti-phase intensity noises defined as
\begin{align}
\widetilde{\delta F}_{\mathrm{In}}(f)&=\frac{1}{\sqrt{2}}\left[\widetilde{\delta F_x}(f)+\widetilde{\delta F_y}(f)\right]\ ,\label{eq20}\\
\widetilde{\delta F}_{\mathrm{Anti}}(f)&=\frac{1}{\sqrt{2}}\left[\widetilde{\delta F_x}(f)-\widetilde{\delta F_y}(f)\right]\ .\label{eq21}
\end{align}
\begin{figure}[]
\centering
\includegraphics[width=0.8\columnwidth]{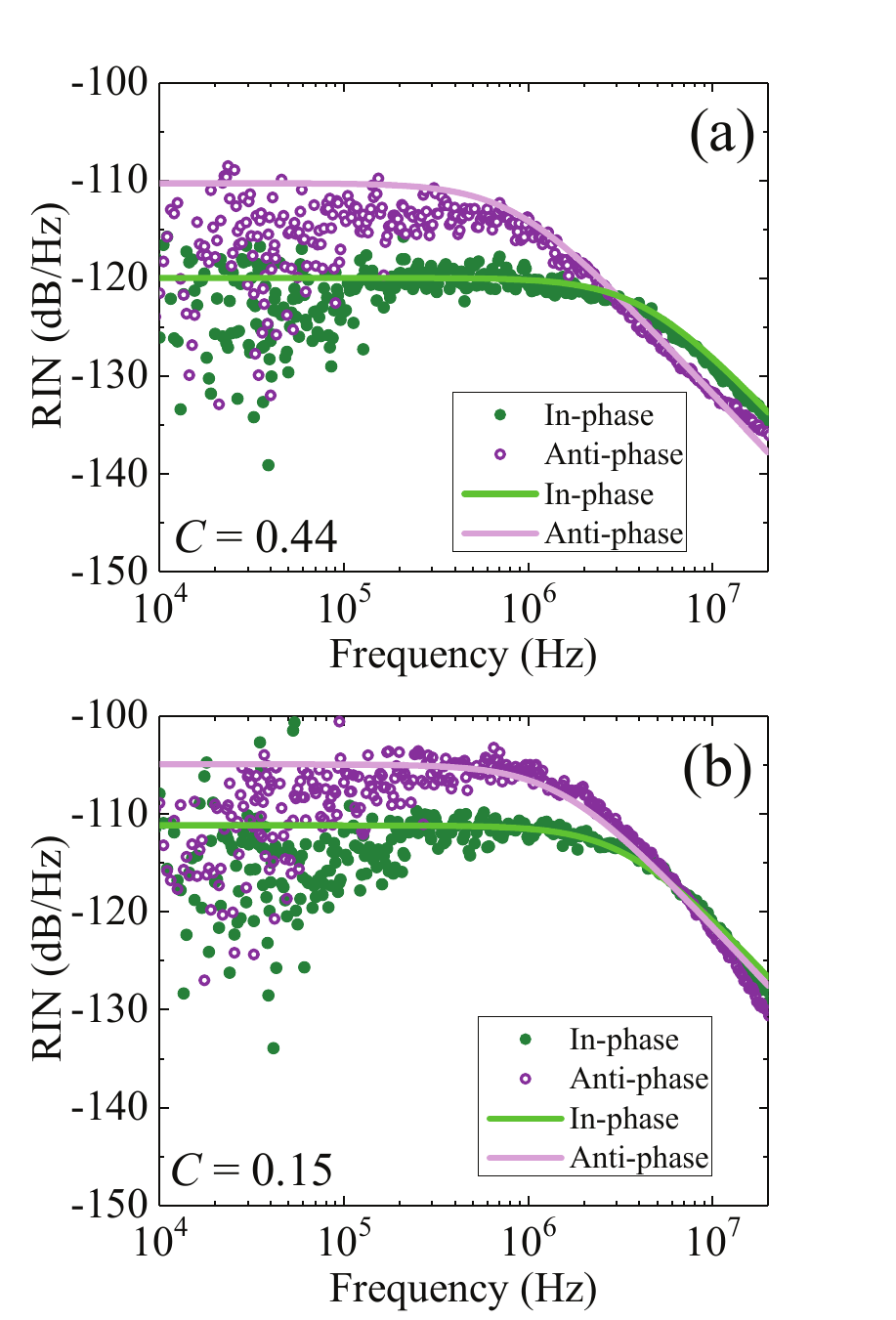}
\caption{Spectra of the in-phase and anti-phase relative intensity noise mechanisms. (a) and (b) correspond to cavity lengths equal to 48.0 mm and 49.5 mm, respectively. The theoretical spectra are obtained with the same parameters as in Fig.\ \ref{Fig07}.}
\label{Fig09}
\end{figure}

One can deduce the spectra of $\widetilde{\delta F_{\mathrm{In}}}(f)$ and $\widetilde{\delta F_{\mathrm{Anti}}}(f)$ from those of the two laser modes, as shown in  Fig.\ \ref{Fig09}. One notices that the anti-phase noise is dominant over the in-phase noise at low frequencies, and conversely at higher frequencies. By combining Eq. (\ref{eq17}) with Eqs. (\ref{eq20}) and (\ref{eq21}), the correlation between $\widetilde{\delta F_x}(f)$ and $\widetilde{\delta F_y}(f)$ reads
\begin{align}\Theta_{\delta F_x-\delta F_y}(f)=&\frac{\left\langle\left|\widetilde{\delta F}_{\mathrm{In}}(f)\right|^2-\left|\widetilde{\delta F}_{\mathrm{Anti}}(f)\right|^2\right\rangle}{2\sqrt{\left\langle\left|\widetilde{\delta F_x}(f)\right|^2\right\rangle\left\langle\left|\widetilde{\delta F_y}(f)\right|^2\right\rangle}}\nonumber\\
&+\frac{\left\langle\widetilde{\delta F}_{\mathrm{In}}^*(f)\widetilde{\delta F}_{\mathrm{Anti}}(f)-\widetilde{\delta F}_{\mathrm{In}}(f)\widetilde{\delta F}_{\mathrm{Anti}}^*(f)\right\rangle}{2\sqrt{\left\langle\left|\widetilde{\delta F_x}(f)\right|^2\right\rangle\left\langle\left|\widetilde{\delta F_y}(f)\right|^2\right\rangle}}\ .\label{eq22}
\end{align}
If now we suppose that the two modes have nearly identical gains and losses, the second term can be neglected, leading to:
\begin{equation}
\Theta_{\delta F_x-\delta F_y}(f)\simeq\frac{\left\langle\left|\widetilde{\delta F}_{\mathrm{In}}(f)\right|^2-\left|\widetilde{\delta F}_{\mathrm{Anti}}(f)\right|^2\right\rangle}{2\sqrt{\left\langle\left|\widetilde{\delta F_x}(f)\right|^2\right\rangle\left\langle\left|\widetilde{\delta F_y}(f)\right|^2\right\rangle}}\ .\label{eq23}
\end{equation}
Thanks to this expression, the spectral behavior of the correlation between $\delta F_x$ and $\delta F_y$ can be readily explained. Since the difference between the in-phase and anti-phase intensity noises is large at low frequencies, the correlation amplitude is high. The $\pi$ correlation phase observed in Figs. \ref{Fig08}(b,d) is consistent with the anti-phase mechanism being dominant at low frequencies. However, the spectral bandwidth of the anti-phase mechanism is smaller than the one of the in-phase mechanism, leading to a crossing of the two curves at a frequency of the order of a few MHz. Close to this crossing point, when the difference between the amplitudes of the two mechanisms becomes small, the correlation amplitude decreases, leading to the appearance of a dip in the correlation spectra of Figs. \ref{Fig08}(a,c). At higher frequencies, the correlation phase tends to be $0$ when the in-phase noise dominates. 

Also, from Fig.\ \ref{Fig09}, one can deduce that when the coupling strength $C$ decreases, the antiphase noise amplitude becomes closer to the amplitude of the in-phase noise at low frequencies. This fact indicates that the coupling between the two modes favors antiphase noise with respect to in-phase noise.

%	The correlations between the PNOB and intensity noises are shown in the fig.10. The correlation amplitude is high at low frequency, and began to decrease at around 1MHz, then keep lower at higher frequency. The large discrepancy between the theoretical results and experimental results in the high frequency can be attributed to the fact that the phase noises approach the noise floor when the frequency higher than several MHz. The different correlation amplitude related to the two polarized modes is attributed to the different gains and losses. The correlation phase related the two polarized modes is all ways keep 180° difference. There is a 90° phase shift between the lower frequency and higher frequency , which can be accounted for resulting from the first-order filter effect of the thermal noise. 

\section{Discussion}\label{discussion}
The experimental and theoretical results reported in the preceding Sections show that the intensity and the beatnote phase noise of the DF-VECSEL do not depend only on the pump noise but also on the laser parameters. Moreover, comparison between theory and experiment has allowed us to gain confidence on the validity of our model. The aim of the present section is thus to use this model to discuss the possible paths to minimize these noises.
\subsection{Minimizing the RIN Transfer}
\begin{figure}[]
\centering
\includegraphics[width=0.8\columnwidth]{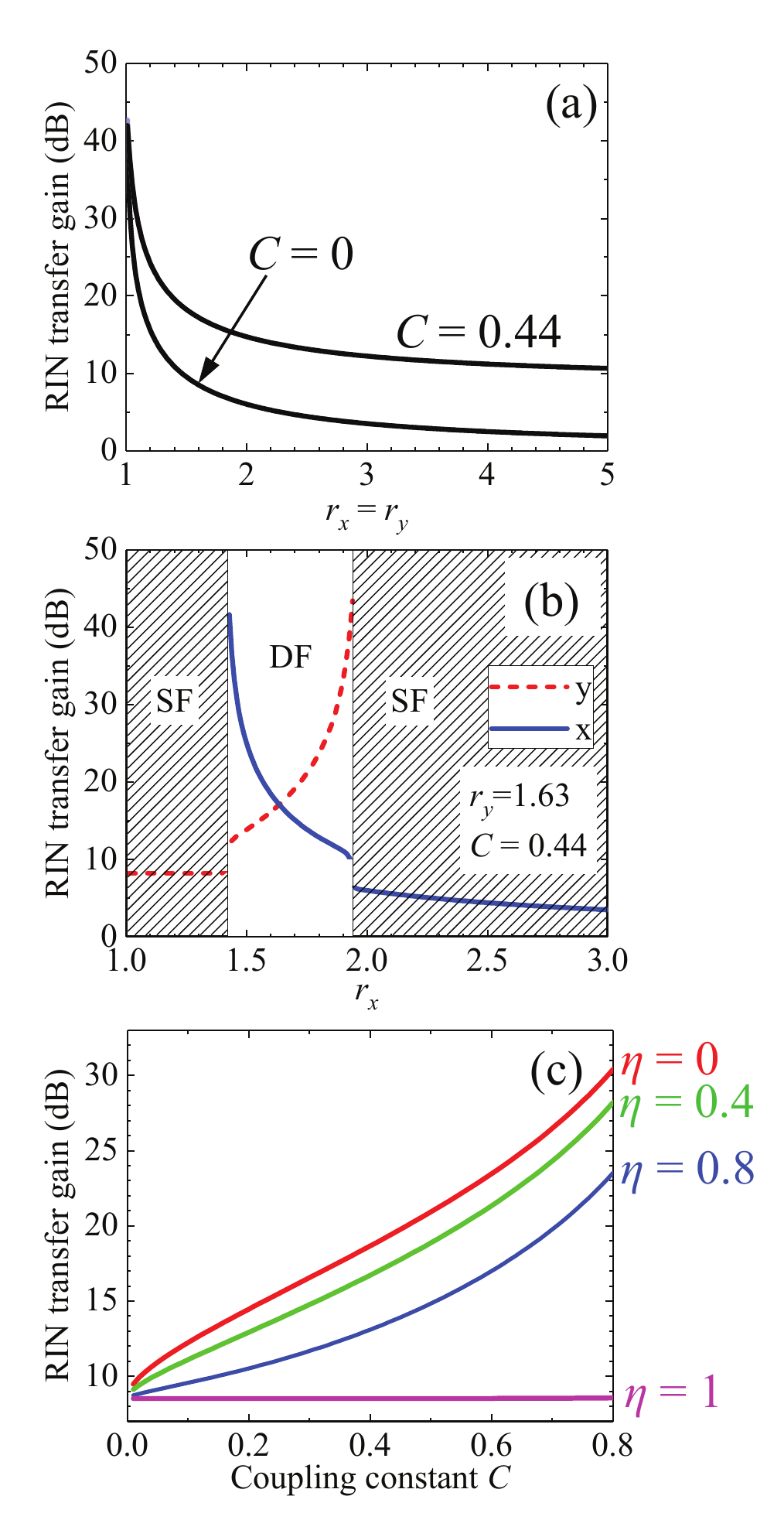}
\caption{(a) Evolution of the pump to DF-VECSEL RIN transfer gain at low frequencies ($f=50\,\mathrm{kHz}$) as a function of the excitation ratio of the laser when the two modes have the same excitation ratios ($r_x=r_y$) for two values of the coupling constant $C$. (b) Evolution of the low-frequency RIN gains for the two modes as a function of the excitation ratio of the $x$ mode for a fixed value of $r_y=1.63$. The two thin vertical lines delimit the region of dual-frequency (DF) oscillation from the regions of single-frequency (SF) oscillation. (c) Same as (a) as a function of the coupling constant $C$ for $r_x=r_y= 1.6$ and different values of the pump noise correlation amplitude $\eta$. All other parameters are $\tau_x=13\,\mathrm{ns}$, $\tau_y=16\,\mathrm{ns}$, $\tau=1\,\mathrm{ns}$, $\eta=0.45$ and $\Psi=0$.}
\label{Fig10}
\end{figure}

Figures \ref{Fig07}(a) and \ref{Fig07}(c) show that the relative intensity noises of the DF-VECSEL modes behave like a low-pass filter with a bandwidth related to the photon lifetime. This means that at low frequencies, much below $1/\tau_x$ and $1/\tau_y$, the relative intensity noises of the laser modes are given by the product of the pump RIN times a transfer gain which is independent of the noise frequency in this frequency range. Figure \ref{Fig10} shows the evolution of this intensity noise transfer gain calculated from Eq. (\ref{eq10}) at frequency $f=50\,\mathrm{kHz}$ in several situations. First of all, Fig. \ref{Fig10}(a) presents the case where the two modes have the same excitation ratio ($r_x=r_y$). In this case, one can see that the RIN transfer gain decreases when the laser excitation ratio increases, just like in a single-frequency class-A laser \cite{Baili2009a}. In particular, when $C=0$, the intensity noises of the two modes are the same as the one of a single-frequency laser. Moreover, one can see from this figure that increasing the nonlinear coupling constant from 0 to 0.44 increases the RIN transfer gain by approximately 10 dB. Furthermore, when $C\neq0$, the two modes have to compete for gain. Then the two modes can oscillate together only when their excitation ratio are relatively similar. This leads to Fig. \ref{Fig10}(b), which shows the evolution of the RIN transfer gain for the two modes when $r_x$ is varied while $r_y$ is fixed. The two vertical lines show the range of values of $r_x$ in which the two modes can oscillate together in a stable manner. Outside this range, the imbalance between the two modes is too strong and competition leads to the fact that the stronger mode kills the weaker one \cite{Sargent1975}. In the simultaneous oscillation region, this figure shows that the RIN transfer gains of the two modes are different as soon as $r_x\neq r_y$, and that the intensity noise of the weaker mode strongly increases with the imbalance between the two modes. Finally, Fig. \ref{Fig10}(c) reproduces the evolution of the RIN transfer gain as a function of the coupling constant $C$ when the two modes are exactly balanced ($r_x=r_y$), for different value of the correlation amplitude $\eta$ between the two pump regions. The intensity noise increases with $C$, except when $\eta$ closely approaches 1. In particular, when $\eta=1$, the intensity noise at low frequency becomes very small. This can be understood by considering Fig. \ref{Fig09}: the intensity noise at low frequencies is dominated by the anti-phase noise mode. Since we consider here that the correlated parts of the two pump noises are always in phase ($\Psi=0$), the fact that $\eta=1$ leads to the fact that the anti-phase noise mode is not excited by the pump noise, leading to a drastic reduction of the low-frequency RIN.

In conclusion of this discussion, we can see that the RIN of the dual-frequency VECSEL can be minimized by i) decreasing the nonlinear coupling constant $C$, ii) balancing the excitation ratios of the two modes as exactly as possible and iii) making the in-phase correlation between the pump noises in the two modes as close to one as possible. 

\subsection{Optimization of the Beatnote Phase Noise}
The performances of the atomic clock based on CPT excited by such a DF-VECSEL will largely depend on the phase noise of the beatnote. Of course, by inserting an electro-optic crystal inside the cavity, one can eventually reduce this noise by phase locking the laser to a reference. But, as can be seen from Figs. \ref{Fig07}(b) and \ref{Fig07}(d), the beatnote phase noise extends to frequencies much larger than what can be easily eliminated with such a phase-locked loop. It is thus important to reduce the contribution of the phase noise coming from the pump laser. 
\begin{figure}[]
\centering
\includegraphics[width=0.7\columnwidth]{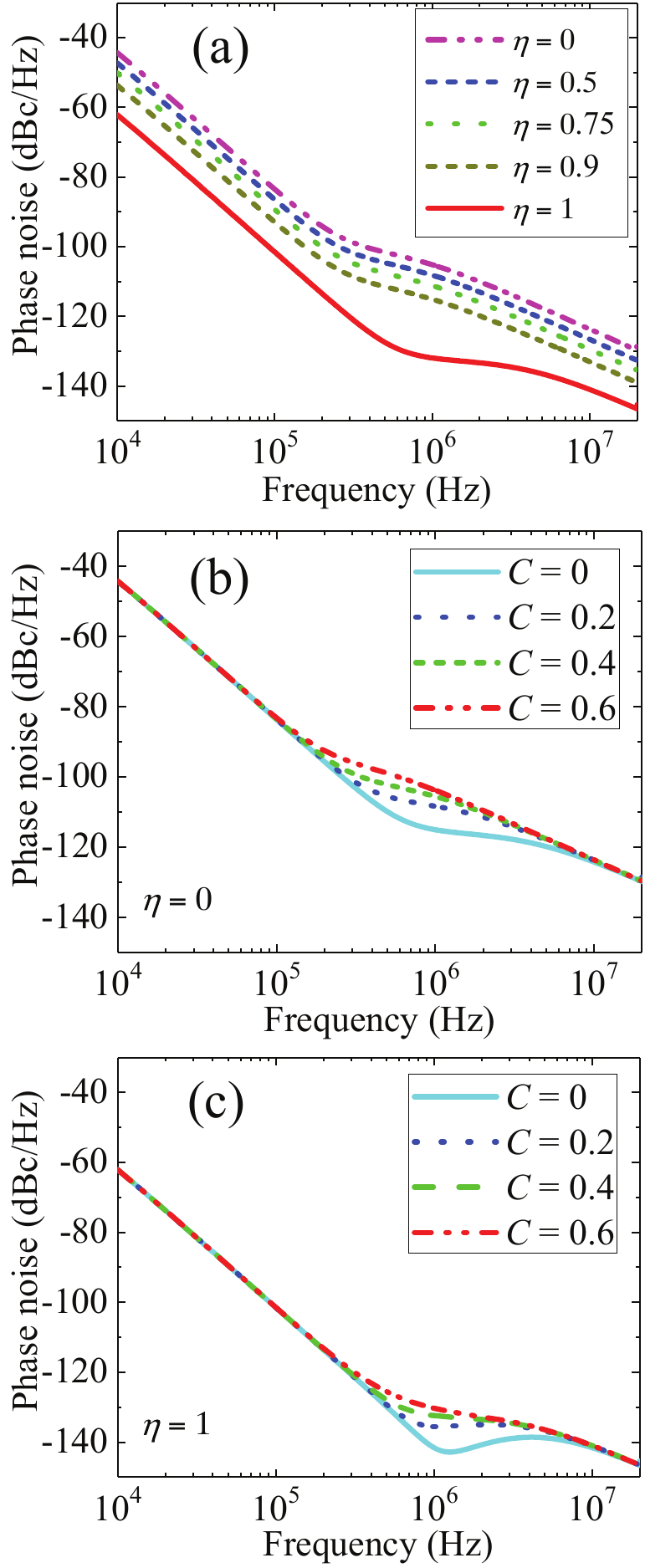}
\caption{Influence of $\eta$ and $C$ on the beatnote phase noise spectra. (a) Evolution of the phase noise spectrum with $\eta$ for $C=0.44$. (b,c) Evolution of the phase noise spectrum with $C$ for (b) $\eta=0$ and (c) $\eta=1$. The values of the other parameters are  $\tau = 1\,\mathrm{ns}$, $\tau_x = 13\,\mathrm{ns}$, $\tau_y = 16\,\mathrm{ns}$, $r_x = 1.59$, $r_y = 1.63$, $\Psi=0$, $\alpha = 5.2$, , $P_{\mathrm{p},x} = 0.42\,\mathrm{W}$, $P_{\mathrm{p},y} = 0.35\,\mathrm{W}$, $\Gamma_\mathrm{T} = 1.34 \times 10^{-7}\,\mathrm{K}^{-1}$, $\tau_\mathrm{T} = 45\,\mu\mathrm{s}$, $R_{\mathrm{T}} = 40\,\mathrm{K.W}^{-1}$, and $\mathrm{RIN}_{\mathrm{p}}= -133\,\mathrm{dB/Hz}$.}
\label{Fig11}
\end{figure}

Figure \ref{Fig11} shows the influence on the beatnote phase noise of the coupling constant $C$ between the modes and the correlation $\eta$ between the two pumps. It is clear from Fig. \ref{Fig11}(a) that the phase noise decreases when $\eta$ increases, but that this decrease is particularly significant when $\eta$ approaches 1. Figures \ref{Fig11}(b) and \ref{Fig11}(c) show that $C$ has a less striking influence on the beatnote phase noise: increasing $C$ decreases the phase noise only in the vicinity of the transition between the two phase noise mechanisms. But it is clear by comparing Figs. \ref{Fig11}(b) and \ref{Fig11}(c) that the most efficient way to reduce the phase noise is to approach $\eta=1$ with $\Psi=0$. 

\begin{figure}[]
\centering
\includegraphics[width=0.7\columnwidth]{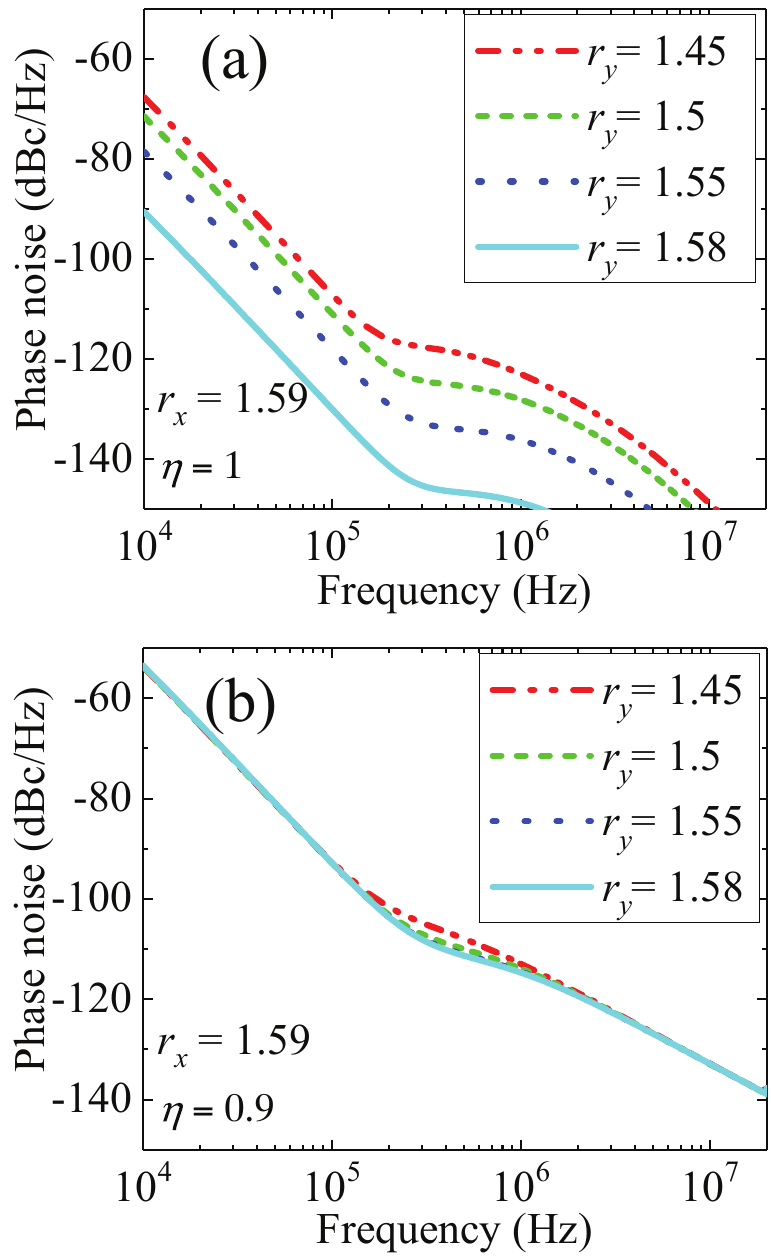}
\caption{Influence of the imbalance between the two modes on the beatnote phase noise spectra. (a) Evolution of the phase noise spectrum with $r_y$ for $r_x=1.59$ and $\eta=1$. (b) Evolution of the phase noise spectrum with $r_y$ for $r_x=1.59$ and $\eta=0.9$. The values of the other parameters are  $\tau = 1\,\mathrm{ns}$, $\tau_x=\tau_y = 13\,\mathrm{ns}$, $r_x = 1.59$, $\Psi=0$, $\alpha = 5.2$,  $\Gamma_\mathrm{T} = 1.34 \times 10^{-7}\,\mathrm{K}^{-1}$, $\tau_\mathrm{T} = 45\,\mu\mathrm{s}$, $R_{\mathrm{T}} = 40\,\mathrm{K.W}^{-1}$, and $\mathrm{RIN}_{\mathrm{p}}= -133\,\mathrm{dB/Hz}$.}
\label{Fig12}
\end{figure}

In the case where $\eta$ is large, Fig. \ref{Fig12} shows the influence of an imbalance between the two modes on the resulting beatnote phase noise. It is clear from Fig. \ref{Fig12}(a) that when $\eta=1$, the last remaining limitation to reduce the beatnote phase noise is the imbalance between the modes. On the contrary, Fig.  \ref{Fig12}(b) shows that for other values of $\eta$, this imbalance plays a marginal role.

To summarize, the minimization of the beatnote phase noise relies on i) minimizing $C$, and ii) making $\eta$ as close to 1 as possible. Finally, the power imbalance between the modes becomes a significant parameter only when $\eta$ gets very close to 1.

\section{Conclusion}
In conclusion, we have theoretically and experimentally studied the noise mechanisms in a DF-VECSEL developed to probe a CPT cesium clock. More precisely, we have analyzed how the pump intensity noise propagates, through the laser dynamics, to the intensity noises of the two modes and to the phase noise of the beatnote between the two modes. In particular, by measuring the parameters of the pump noise and the intensity and beatnote phase noises of the DF-VECSEL, we have obtained a good agreement with our model based on generalized rate equations. This has allowed us to isolate the parameters that play an important role on the DF-VECSEL noises. Beyond the trivial conclusion that the pump noise should be made as small as possible, it appears that the same parameters must be optimized to reduce both the intensity noise and the beatnote phase noise. First, the excitation ratios of the two modes must be as similar as possible. This can be performed by carefully aligning the laser and adjusting the overlap between the two modes and the region of the semiconductor structure that is pumped by the pump laser. Second, the coupling constant between the two modes must be made as small as possible. This means that cross-saturation must be minimized, for example by increasing the spatial separation between the two modes in the active medium. Finally, the correlation between the noises of the two regions of the pump that pump the two modes must be as close as possible to one, and in phase. This condition is somewhat contradictory with the fact that $C$ must be made as small as possible. However, one could probably satisfy these two criteria by splitting the pump into two identical pump beams and pumping the two spatially separated modes with pump beams exhibiting exactly the same noises. The problem then is to find a low-noise pump laser that exhibits enough output power to pump two well separated modes to allow them to oscillate far enough above threshold to have a stable laser emission regime. This is the subject of our present investigations.

% or


%\appendices
%% you can choose not to have a title for an appendix
%% if you want by leaving the argument blank
%\section{}

\section*{Acknowledgment}
The authors thank Gr\'egoire Pillet for providing the computer codes for computing the noise spectra, Syamsundar De for his help, and Gaelle Lucas-Leclin and Sylvie Janicot. This work was supported by the Agence Nationale de la Recherche (grant number ANR-15-CE24-0010-04) and the Direction G\'en\'erale de l'Armement. The work of HL, FB, FG, GG and GB is performed in the framekwork of the joint lab between Laboratoire Aim\'e Cotton and Thales Research \& Technology. 
\ifCLASSOPTIONcaptionsoff
  \newpage
\fi

\end{document}